\documentclass{ws-procs975x65}
\usepackage{graphicx}
\usepackage{latexsym}
\usepackage{amssymb}

\begin{document}

\title{Killing horizons,  throats and bottlenecks in the ergoregion of the Kerr spacetime}
\author{D. Pugliese and H. Quevedo}
\address{
Institute of Physics, Faculty of Philosophy \& Science,
  Silesian University in Opava,
 Bezru\v{c}ovo n\'{a}m\v{e}st\'{i} 13, CZ-74601 Opava, Czech Republic\\
          Dipartimento di Fisica, Universit\`a di Roma ``La Sapienza", I-00185 Roma, Italy Instituto de Ciencias Nucleares, Universidad Nacional Aut\'onoma de M\'exico,  AP 70543, M\'exico, DF 04510, Mexico
          \\
          E-mail:d.pugliese.physics@gmail.com}
\date{\today}
\begin{abstract}
The properties of Kerr black holes (\textbf{BHs})and naked singularities (\textbf{NSs})
are investigated by using stationary observers and their limiting frequencies.
We introduce the concept of \textbf{NS} Killing throats and bottlenecks
for slowly spinning \textbf{NSs} to describe the frequency of stationary observers.
In particular, we show the frequency on the horizon can be used to point out a connection between \textbf{BHs} and \textbf{NSs}
and to interpret the horizon in terms of frequencies. The analysis is performed on the equatorial plane of the ergoregion.
\end{abstract}

\bodymatter
\def\be{\begin{equation}}
\def\ee{\end{equation}}
\def\bea{\begin{eqnarray}}
\def\eea{\end{eqnarray}}
\newcommand{\bt}[1]{\mathbf{\mathtt{#1}}}
\newcommand{\tb}[1]{\textbf{{{#1}}}}
\newcommand{\rtb}[1]{\textcolor[rgb]{1.00,0.00,0.00}{\tb{#1}}}
\newcommand{\btb}[1]{\textcolor[rgb]{0.00,0.00,1.00}{\tb{#1}}}
\newcommand{\otb}[1]{\textcolor[rgb]{1.00,0.50,0.00}{\tb{#1}}}
\newcommand{\gtb}[1]{\textcolor[rgb]{0.00,.50,0.00}{\tb{#1}}}
\newcommand{\ptb}[1]{\textcolor[rgb]{0.70,0.00,0.70}{\tb{#1}}}
\newcommand{\il}{~}
\newcommand{\Qa}{\mathcal{Q}}
\section{Introduction}
We  study  the orbital angular frequencies of  stationary observers  in the Kerr spacetime. We introduce the concept of
 Killing throats that arise  in the spacetime of \textbf{NSs} and can be interpreted as the ``opening" and disappearance of Killing horizons. Killing bottlenecks are identified as ``restrictions'' of Killing throats that appear in the case of weak naked singularities\footnote{{The concept of strong and weak \textbf{NSs}, defined through the values of the spin parameter, has been explored in several works\cite{Kerr,Kerr-Newman,ergon,observers}. However, they can also be defined as strong curvature singularities \cite{strong}.}}
(\textbf{WNSs}) for which the spin-mass ratio is close to the value of the extreme \textbf{BH}.
To explore these \textbf{NS} effects, and  considering  the dynamics of the zero angular momentum observers (\textbf{ZAMOs}), we introduce
the concept of  ``metric bundles" and ``extended planes". In this work, we limit  the analysis of the equatorial plane of the  Kerr spacetime.
The generalization to the case of the Reissner-Nordstr\"om and Kerr-Newmann spacetimes is presented elsewhere \cite{observers}.

We first analyze  the behavior of the frequency of a stationary observer in terms of the radial distance and the spin parameter
$a$ of the source. In this way, we find in the \textbf{NS} region a particular set of curves  that we identify as the Killing throat.
In the case of \textbf{WNSs}, for which $a/M\in ]1,2]$, the Killing throats show ``restrictions''  identified as Killing bottlenecks. To explore the properties of the bottlenecks, we introduce the concept of extended plane which is a graph relating a particular characteristic of a spacetime in terms of the parameters entering the corresponding spacetime metric. Any curve on the extended plane represents, therefore, a metric bundle, i.e., {a family of spacetimes  defined by a characteristic photon orbital frequency $\omega$  and  characterized by a particular relation between the metrics  parameters.}
In the case of the Kerr spacetime, the extended plane turns out to establish a relation between \textbf{BHs} and \textbf{NSs}.
As a consequence,  \textbf{WNSs} turn out to be related to the appearance of {(a portion of)} the inner horizon, whereas strong naked
singularities \textbf(\textbf{SNSs}) with $a>2M$ are related to the outer  horizon.

This work is organized as follows.
In Sec. (\ref{Sec:Stationar}), we discuss the concept of stationary observers, introducing the concepts of  Killing throats and Killing bottlenecks.
A discussion  of the significance of the Killing bottlenecks and  their  possible origin  is presented in Sec.\il(\ref{Sec:p-lS}).
In Sec.(\ref{Sec:extende}), we introduce a possible generalization of the Killing horizon definition in the extended plane. Finally,
we discuss a possible connection between black holes and naked singularities,
revisiting the definition of horizons and the role played by \textbf{NSs} in horizon  construction.
Final remarks follow in Sec.\il(\ref{Sec:remark}).

\section{Stationary observers and light surfaces}
\label{Sec:Stationar}
Consider the Kerr spacetime in Boyer-Lindquist (BL) coordinates, $(t, r, \phi, \theta)$, with $M\geq0$ as the mass parameter and
$a\equiv J/M\geq0$ as the  \emph{specific} angular momentum or \emph{spin}, where $J$ is the \emph{total} angular momentum of the gravitational source. The horizons and ergospheres radii are given by $r_\pm = M\pm \sqrt{M^2 -a^2}$ and
$r_\epsilon ^\pm = M\pm \sqrt{M^2-a^2\cos^2\theta}$.
\emph{Stationary observers} are characterized by a four-velocity of the form
\be
\label{Eq:spectrum}
u^\alpha=\gamma(\xi_t^\alpha+\omega \xi_\phi^\alpha), \quad
\gamma^{-2}\equiv-\kappa(\omega^2 g_{\phi\phi}+2\omega g_{t\phi}+g_{tt}) ,
\ee
where $\gamma$ is a normalization factor with   $\kappa = - g_{\alpha\beta} u^\alpha u^\beta$,
$\xi_{\phi}$  is the rotational Killing field, $\xi_{t}$ is the time-translational Killing field,
and  $\omega$   is a  uniform \emph{angular velocity}
(dimensionless quantity)\footnote{The particular case  $\omega=0$ defines {\em static observers}; these   observers  cannot exist in the ergoregion.}.
In BL coordinates, in which the metric tensor depends on $(r,\theta)$ only;
this means that  $r$ and $\theta$ are constants along  the worldline of each stationary observer (the  observer does not see the spacetime changing along the trajectory). The spacetime  causal structure of the Kerr spacetime can be also  studied by  considering    stationary  observers  \cite{malament}.

Timelike stationary  particles have orbital frequencies in the range
\bea\label{Eq:ex-ce}
\omega\in]\omega_-,\omega_+[ \quad\mbox{where}\quad \omega_{\pm}\equiv \omega_{Z}\pm\sqrt{\omega_{Z}^2-\omega _*^2},
\quad
\omega _*^2\equiv \frac{g_{tt}}{g_{\phi \phi}}=\frac{g^{tt}}{g^{\phi\phi}},\quad \omega_{Z}\equiv-\frac{g_{\phi t}}{g_{\phi\phi}},
\eea
where $\omega_{Z}$ is the orbital angular frequency of the zero angular momentum observers
(ZAMOS) and
$\omega_{\pm}$ are the limiting frequencies of photons orbits, which are solutions of the equation
$ g_{\alpha\beta}\mathcal{L}^\alpha_{\pm}\mathcal{L}^\beta_{\pm}=0$
with $\mathcal{L}_{\pm}\equiv \xi_{t}+\omega_{\pm}\xi_{\phi}$. Killing vectors $\mathcal{L}_{\pm} $ are  generators of Killing  horizons. The Killing vector $\xi_t+\omega\xi_{\phi}$ becomes null at $ r = r_+$, defining the frequency $\omega_+(r_+)=\omega_h$.

\begin{figure*}[ht!]
\begin{tabular}{cc}
\includegraphics[scale=.5]{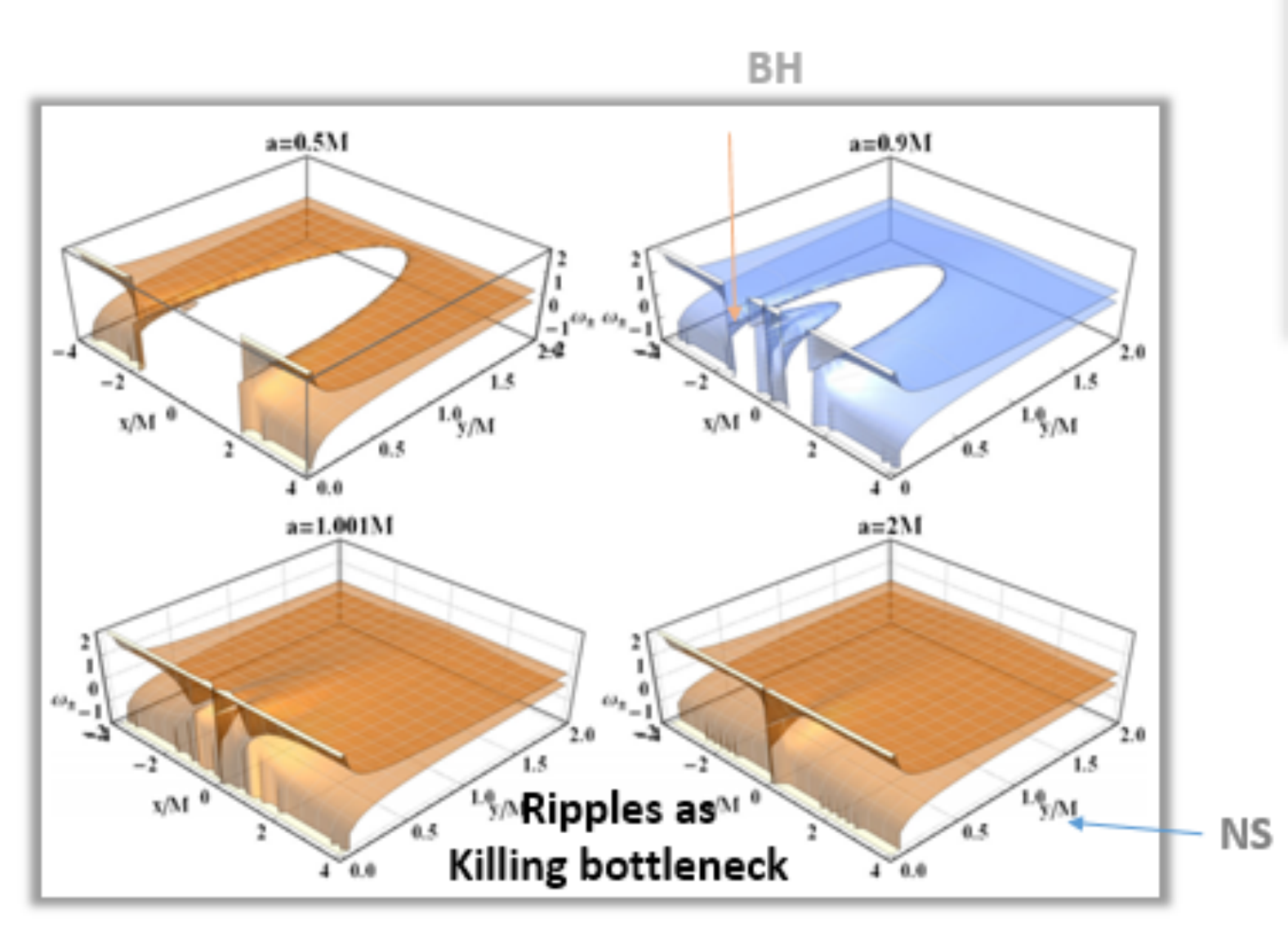}
\includegraphics[scale=.23]{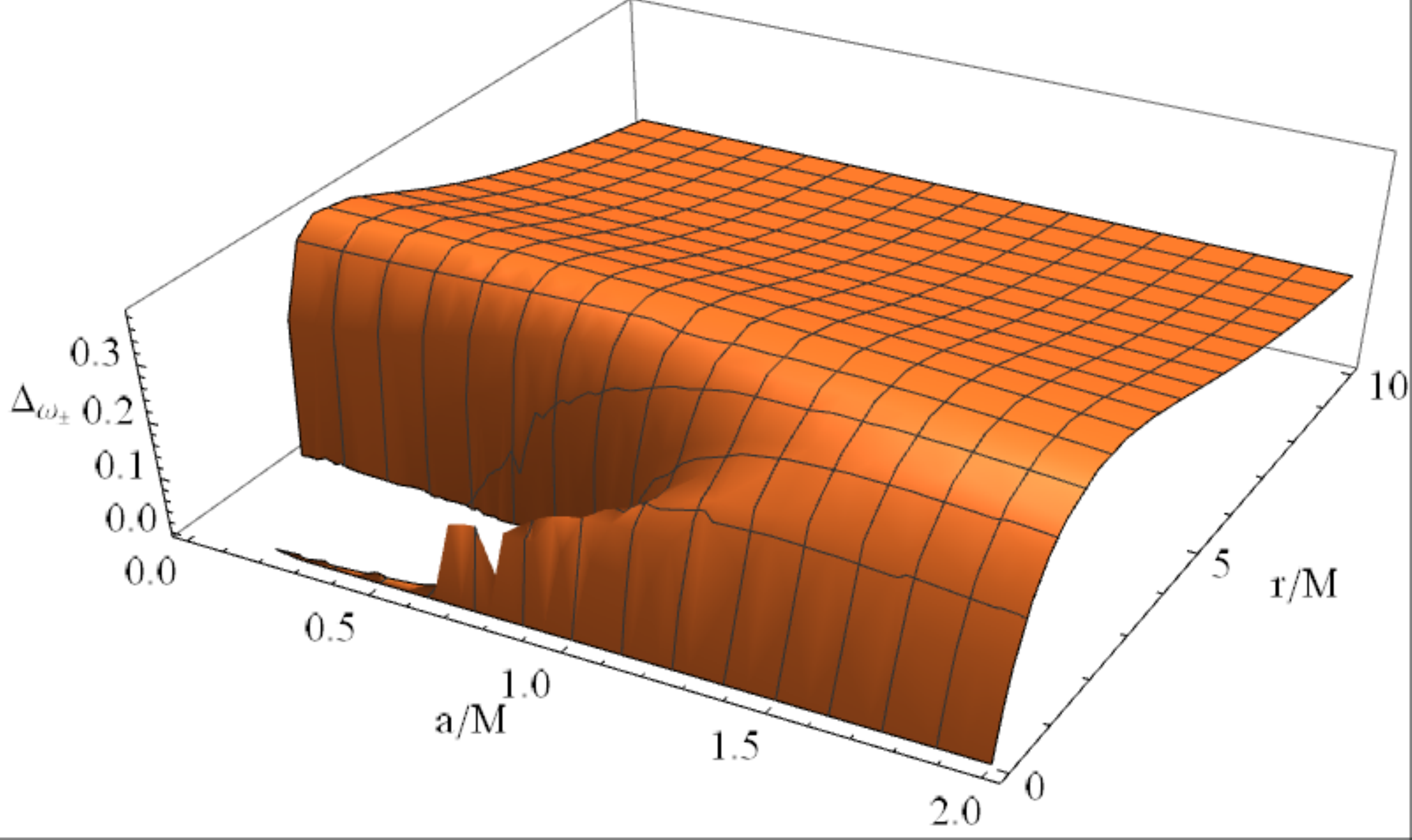}
\end{tabular}
\caption[font={footnotesize,it}]{Left panel: Plots of the frequency surfaces $\omega_{\pm}$
as functions of the radial
distance $r$ in Cartesian coordinates $(x,y)$ for different spin values $a$,
including \textbf{BH}s (upper plots) and \textbf{NS}s  (bottom plots).
Timelike stationary observers are defined in the region bounded by these planes.
In the \textbf{BH} case, the horizons are clearly identified; as the spin increases, the horizons merge and,
in the \textbf{NS} region, a rippled configuration (bottleneck)  appears specially in  the case of \textbf{WNSs} with
($a\in ]M, 2M]) $.
Right panel: Plot of the frequency interval $\Delta \omega_{\pm}=\omega_+-\omega_-$ as a
function of the radius $r/M$ and the spin $a/M$.
The extrema $r_{\Delta}^{\pm}$ and $r^{\pm}_{\blacksquare}$ are solutions of  the equations
$\partial_r\Delta \omega_{\pm}=0$
and $\partial_{a}\Delta \omega_{\pm}=0$, respectively. Bottlenecks are  shown explicitly \cite{observers}.}
\label{Fig:QPlot1}
\end{figure*}

On the equatorial plane $\theta=\pi/2$ of the Kerr spacetime we have that
\be
\label{Eq:comb}
\omega_{\pm}\equiv\frac{2 aM^2\pm M\sqrt{r^2 \Delta}}{r^3+a^2 (2M+r)},\quad  \Delta\equiv r^2-2Mr+a^2
\ee
with the asymptotic behavior
\be
\lim_{r\rightarrow\infty}\omega_{\pm}=0,\quad \lim_{a\rightarrow\infty}\omega_{\pm}=0,\quad
\lim_{r\rightarrow0}\omega_{\pm} \equiv  \omega_0=\frac{M}{a}
\ee
and the particular values
\be
\omega_{h}\equiv\omega_{\pm}(r_+)= \omega_{Z}(r_+)=\frac{a}{2 r_+} ,  \quad
\omega_{\epsilon}\equiv\omega_{+}(r_{\epsilon}^+)=\frac{aM}{2M^2+a^2} .
\ee
We can see that $\omega_-<0$ for $r>r_{\epsilon}^+$, and $\omega_->0$ inside the ergoregion, while $\omega_+>0$  everywhere.
Moreover, since $\omega_+=\omega_-$ on the horizon, stationary observers cannot exist inside this surface. Therefore, $\omega_{\pm}$ are limiting angular velocities for physical observers.
The behavior of the frequencies $\omega_\pm$ is depicted in Fig. \ref{Fig:QPlot1}.
Notice that instead of the spin parameter $a$, we could use the frequency $\omega_0$ or $\omega_h$ to parameterize the Kerr spacetime by using the  relations
\be
a= \frac{M}{\omega_0}  \quad \hbox{or} \quad a = \frac{4M\omega_h}{1+4\omega_h^2}\ ,
\label{omh}
\ee
respectively, which can be prove to be  $1-1$ relations. 
This simple observation has very important consequences. In fact, if we could measure all the values of $\omega_0\in]0,\infty[$ or $\omega_h\in]0,1/2]$, the entire Kerr family of spacetimes would be described by these frequencies. From a practical point of view, however, we can expect that the frequency on the (outer) horizon $\omega_h$ is a more suitable candidate for being measured. This, again, is an interesting fact that seems to be related to the well-known property that all the physical degrees of freedom of a black hole are encoded on the horizon.

The angular frequency $\omega_0$ (``frequency of the singularity")  emerges as a relevant quantity in this analysis and, as it is clear from Fig.\il\ref{Fig:QPlot1}, it can be used to differentiate  between low spin
$a\in[0,M]$ sources, where the frequency gap  $\Delta \omega_{\pm}=\omega_+-\omega_-$ is larger, and  \textbf{WNSs} and \textbf{SNSs}, where stationary observers are not defined (asymptotically, for large $a/M$, $\omega_0\approx0$ and $\Delta \omega_{\pm}\approx0$ for a fixed orbit $r$).

The constant $\omega_h$ is also important in the context of \textbf{BH} thermodynamics. Indeed, from the fundamental Bekenstein-Hawking
entropy equation, $S=\frac{1}{4}A_h$, where $A_h=A_h(r_+)$ is the area of the outer horizon, one can derive all the thermodynamic properties of a \textbf{BH}, including its temperature (surface gravity) and angular velocity at the horizon\footnote{In fact,
 one can write the Hawking temperature as
 $T_{H}= {\hbar c\kappa }/{2\pi k_{B}}$, where $k_{B}$ is the Boltzmann constant and $\kappa$ is the surface gravity. %
Temperature $T= \kappa/(2\pi)$;
entropy $S= A/(4\hbar G)$, where $A =$ area of the horizon $A = 8\pi mr_+$;
pressure $p= - \omega_h $;
volume $V= G  J/c^2$ ($J = amc^3/G$);
internal energy $U$= GM ($M = c^2m/G $= mass) and  $m$ is the mass.}.
 {The frequency $\omega_h$ also determines the  uniform (rigid) angular velocity on the horizon, representing the fact that the \textbf{BH} rotates rigidly.
This quantity enters directly into the definition of the \textbf{BH} surface gravity and, consequently, into the formulation of the rigidity theorem.
\textbf{BH} thermodynamics is established  through the  quantities $(r_+,\omega_h)$ and their variation (in $r>r_+$);
consequently, from the point of view of the geometric laws of the  \textbf{BH} thermodynamics considered here,
the universe in the outer region, $r>r_+$, is  regulated exclusively by the outer  horizon  $r_+
$ (the region ``inside'' the black hole, i.e., $r<r_+$, does not directly enter in the outer region thermodynamics).
This aspect will have a correspondence in the concept of inner horizon confinement (in the extended plane) that we will
show in Sec.\il(\ref{Sec:extende}).}
{The \textbf{BH} surface gravity is defined from  the variation  of the Killing field $\mathcal{L}$ norm on the outer \textbf{BH} horizon.  In this analysis we study this norm and its variation on the equatorial plane of the Kerr geometries (i.e. $r>0$ and $\theta=\pi/2$).
We focus   particularly on the   ergoregion as     the outer boundary of the ergosurface ($r_{\epsilon}^+=2M$) is  a spin-independent  quantity (in the static limiting metric,  where $r_{\epsilon}^+=2M$, there is an event horizon associated to the Killing
$\xi_t$--apparent, trapped, Killing and events  horizons in stationary and static spacetimes coincide.). The relevance of this special feature of the outer ergosurface on the equatorial plane consists in the  variation of the norm of the Killing field $\mathcal{L}$  with the dimensionless spacetime spin $a/M$.
The off-equatorial case is instead considered in \cite{renmants}.}

Since $S$ depends only on the radius of the outer horizon, which in turn depends only on $M$ and $a$, we can replace the spin parameter $a$ by the frequency parameter $\omega_h$, by using the $1-1$ relation (\ref{omh}). This means that $\omega_h$ can play the role of a thermodynamic variable for a \textbf{BH}. It would be interesting to investigate how the use of the frequency $\omega_h$, instead of the angular momentum $J=aM$, as a thermodynamic variable would affect the thermodynamics of the corresponding \textbf{BHs}.

\textbf{
{Killing vectors and Killing horizons}}

It is convenient here to  review some well-known facts about Killing horizons
and Killing vectors.
A \textbf{Killing horizon} is a \emph{null} surface, $\mathcal{S}_0$, whose \emph{null} generators coincide with the orbits of an
one--parameter group of isometries (i. e., there is a Killing field $\mathcal{L}$
which is
normal
to $\mathcal{S}_0$). Therefore, it  is  a lightlike hypersurface (generated by the flow of a Killing vector) on
which the norm of a Killing vector goes to zero. In static  \textbf{BH} spacetimes, the
event, apparent, and Killing horizons  with respect to the  Killing field   $\xi_t$ coincide.
In the Schwarzschild spacetime, therefore, $r=2M$ is the  Killing horizon with \emph{respect} to the  Killing vector
$\partial_t$.  The event horizons  of a spinning \textbf{BH}  are   Killing horizons   with respect to  the Killing field $\mathcal{L}_h=\partial_t +\omega_h \partial_{\phi}$, where  $\omega_h$ is  defined as the angular velocity of the horizon
(the event horizon of a stationary black hole must be a Killing horizon).
In \cite{renmants}, we also consider the case of the Reissner--Nordstr\"om and Kerr-Newman spacetimes.

\textbf{ Killing vectors and BH surface gravity}

The \textbf{surface gravity} of a \textbf{BH} may be defined as the  \emph{rate} at which the norm of the Killing vector vanishes from  outside.
The surface gravity for the Kerr \textbf{BH} metric, $\mathcal{SG}_{Kerr}= (r_+-r_-)/2(r_+^2+a^2)$, is a conformal invariant of the metric, but it re-scales with the conformal Killing vector.
Therefore, it  is not the same on all generators (but obviously it is constant along one specific generator because of the symmetries).

\textbf{Surface gravity and frequencies}

 The  Kerr  \textbf{BH} surface gravity  can be decomposed as  $\kappa =\kappa_s-\gamma_a$, where $\kappa_s\equiv { {1}/{4M}}$ is the Schwarzschild  \textbf{BH} surface gravity, while  $\gamma_a=M\omega_{h}^{2}$   is the contribution due to the additional component of the
\textbf{BH} intrinsic spin; $\omega_{h}$ is, therefore, the  angular velocity (in units of $1/M$) on the \emph{event horizon}.
The (strong) rigidity theorem  connects then the event horizon with a Killing  horizon, stating  that,  under suitable conditions,   the event horizon of a stationary (asymptotically flat) solution
 satisfying suitable hyperbolic equations is a Killing horizon.
The surface area of the \textbf{BH} event horizon
is non-decreasing with time (which is the content of the second   law of black hole thermodynamics).
 The \textbf{BH} event horizon of the
stationary  solution
is a { Killing horizon
with constant surface gravity (zeroth  law)}\footnote{Progenitors, as stars or galaxies, have generally  spin  $a = J/(Mc) $  usually bigger than their mass $m = GM/c^2$.  During the  gravitational collapse, the  body should  lose
mass and   angular momentum,  generating the Killing horizons, ending  up as a black hole. 
Penrose  Cosmic
Censorship Hypothesis constrains the gravitational collapse  from  good (physically realistic) initial conditions for the progenitors to end up in a \textbf{BH}; this result, however,
is still an hypothesis, strictly depending on the ``physical" initial conditions}.
Thus $\Lambda=\mathcal{L}^\alpha\mathcal{L}_\alpha $ is  constant on the horizon.
The surface gravity is then defined as the constant $\kappa: \nabla^\alpha\Lambda=-2\kappa \mathcal{L}^\alpha$
 (on the outer horizon $r_+$). Alternatively, it holds that
  $\mathcal{L}^\beta\nabla_\alpha \mathcal{L}_\beta=-\kappa \mathcal{L}_\alpha$ and  $L_{\mathcal{L}}\kappa=0$, where $L_{\mathcal{L}}$ is the Lie derivative (a non affine geodesic equation).
In other words, $\kappa$ is  constant on the orbits of $\mathcal{L}$.

We here consider $\Lambda$ as a function of $(r, a)$ (as we are restricting the analysis to the equatorial plane). To analyze the dependence
on $a$, we introduce the concept of the \emph{extended plane} $\pi^+$ as the set of points $(a/M,\Qa)$, where $\Qa$ is any quantity that characterizes the spacetime and depends on $a$. In general, the extended plane is an
$(n+1)$-dimensional surface, where $n$ is the number of independent parameters that enter $\Qa$  \cite{renmants}.
An example of an extended plane is plotted in Fig.\il\ref{Fig:QPlot1}, which corresponds to the 2D set of points $(a/M,\omega_\pm)$.
The frequency $\omega_{\pm}$ shows ``ripples", which emerge when the spin is within the interval $a\in]M, 2M]$. The surfaces become connected  and the two horizons disappear, leaving  as ``remnants" in the planes the ``Killing bottlenecks".
Those ripples appeared also in different analysis of the stationary observes  \cite{observers}.

\subsubsection{Light surfaces}
\label{Sec:p-lS}

\begin{figure*}[ht!]
\begin{tabular}{ccc}
\includegraphics[scale=.4]{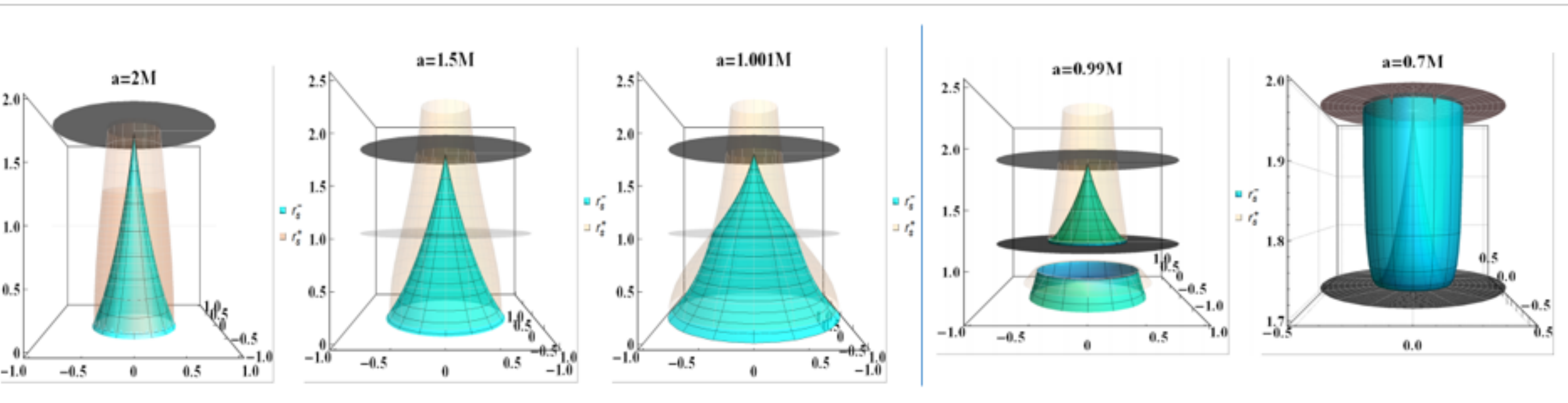}
\end{tabular}
    \caption[font={footnotesize,it}]{{Plots of the surfaces $r_{s}^{\pm}$ (in units of mass) versus the frequency
		$\omega$ for different spin values $a/M$, including \textbf{BH} and \textbf{NS} geometries. The Killing throat and bottleneck are
		plotted \cite{observers}.}}
\label{Fig:QPlot}
\end{figure*}

\begin{figure}[h!]
\centering
\begin{tabular}{lr}
\includegraphics[scale=.45]{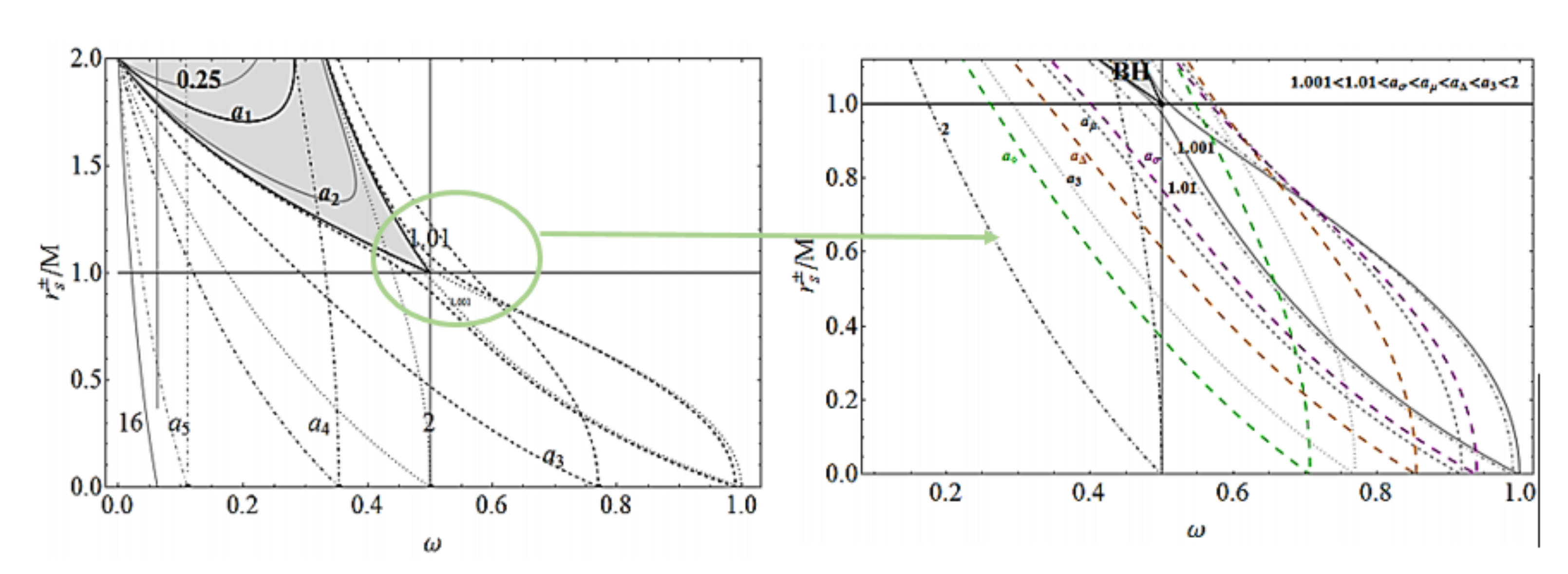}
\end{tabular}
\caption{The radii $r_s^{\pm}$ versus  the frequency $\omega$ for different values of the spin $a/M$
(numbers close to the curves). The gray region is the only region allowed for the case of \textbf{BH} spacetimes.
The surfaces $\hat{r}_{\pm}$ at $a=M$ (extreme-\textbf{BH}-case) are shown in  black-thick (from \cite{observers}).}
\label{Fig:Trav-inr-B}
\end{figure}

Light surfaces are determined by the solutions of the equation
$ g_{\alpha\beta}\mathcal{L}^\alpha_{\pm}\mathcal{L}^\beta_{\pm}=0$, which in the case under consideration can be expressed as
\cite{observers}
\bea
&&\frac{r_s^-}{M} \equiv \frac{2 \beta_1 \sin \left(\frac{1}{3} \arcsin\beta_0\right)}{\sqrt{3}}
,\quad
\frac{r_s^+}{M} \equiv \frac{2 \beta_1 \cos \left(\frac{1}{3}\arccos(-\beta_0)\right)}{\sqrt{3}}
\\
&&
\mbox{where}\quad
\beta_1\equiv\sqrt{\frac{1}{\omega ^2}-\frac{1}{\omega_0^2}},\quad \beta_0\equiv\frac{3 \sqrt{3} \beta_1 \omega ^2}{\left(\frac{\omega }{\omega_0}+1\right)^2}.
\eea
The surfaces $r_s=r_s^+\cup r_s^-$  are represented for \textbf{BHs}  and \textbf{NSs} in Fig.\ref{Fig:QPlot}.
In the \textbf{BH} case, we note the disconnection due to the presence of the horizons, which corresponds to the disconnection in the frequency planes of Fig.\il\ref{Fig:QPlot1}, where we limit the plots to the region $r_s<r_{\epsilon}^+$. Physical (timelike) observers are located between the inner cone and the outer cylinder surfaces of Fig.\il\ref{Fig:QPlot}. As the spin increases into the
\textbf{NS} regime, the surfaces merge at $a\geq M$, giving rise to a connected  surface  $r_s$ ($r_s^+\cup r_s^-$),
the Killing throat (or tunnel). For spins $a\in]0,2M]$, the Killing  bottleneck appears as restriction of the surface
$r_s=r_s^+\cup r_s^-$. We focus the analysis on the Killing  bottleneck considering
a $2D$ representation of the Killing throat versus the frequency $\omega$ at different  spins, close to  values where the bottleneck appears.
Fig.\il\ref{Fig:Trav-inr-B} shows the behavior for  \textbf{BHs} (gray region) and \textbf{NSs}.
In the \textbf{BH} case, we do not consider the  region inside the inner horizon (i.e. $r<r_-$).
The outer horizon is the tangent point $r_+$ to the curve $r_s$ and the minimum (regular) point of the surface
$r_s(\omega)$. For the extreme Kerr spacetime the horizon is a (non regular) cusp point  in  this representation.
Each Killing throat is associated to one spacetime geometry.
For example,   the  Killing throat  defined for $a=2M$ has a characteristic
singularity frequency $\omega_0=M/a=0.5$, which is the limiting frequency at the singularity (note that the restriction of the throat  approaching  the  singularity corresponds to a null  frequency gap $\Delta \omega_{\pm}$, meaning that there are no physical stationary observers close to the singularity).
At a fixed value $r_s(\omega)$, one orbit on a throat $a=2M$, corresponds to two (positive) photon orbital frequencies
$\omega_+$ and  $\omega_-$ (a part  on  the singularity $r_s=0$); viceversa, there can be  two orbits $r_1(\omega)<r_2(\omega)$ on the  surface $r_s$  (vertical lines in Fig.\il\ref{Fig:QPlot}),
 where photons have equal orbital frequency (note that the photon orbital frequency $\omega_{\pm}$ are limiting frequencies for timelike particles; therefore, they  determine also the range of possible values of $\omega$ for the physical stationary observers).

{
Killing bottleneck appears as restriction of the Killing throat in the light surfaces analysis and as  ripples  in the frequency planes of
Fig.\il\ref{Fig:QPlot1}.   We considered the possibility that   these  ``horizons remnants" in \textbf{WNSs} were caused by failure of the \textbf{BL} coordinates  ``close'' to the   $a=M$, in the region near the horizon $r_+=r_-=M$; nevertheless,  Killing bottlenecks  appear in the whole spin range $a\in] M, 2M]$.
 To understand if the Killing bottlenecks are due to some known geometrical  properties of the singularities,  we focus our analysis on the spins   $a\in]M,2M]$. Considering \textbf{ZAMOs}  dynamical properties,  we  highlight  some particular spin values of this range.
}


%
\begin{figure}[h!]
\centering
\begin{tabular}{lr}
\includegraphics[scale=.3]{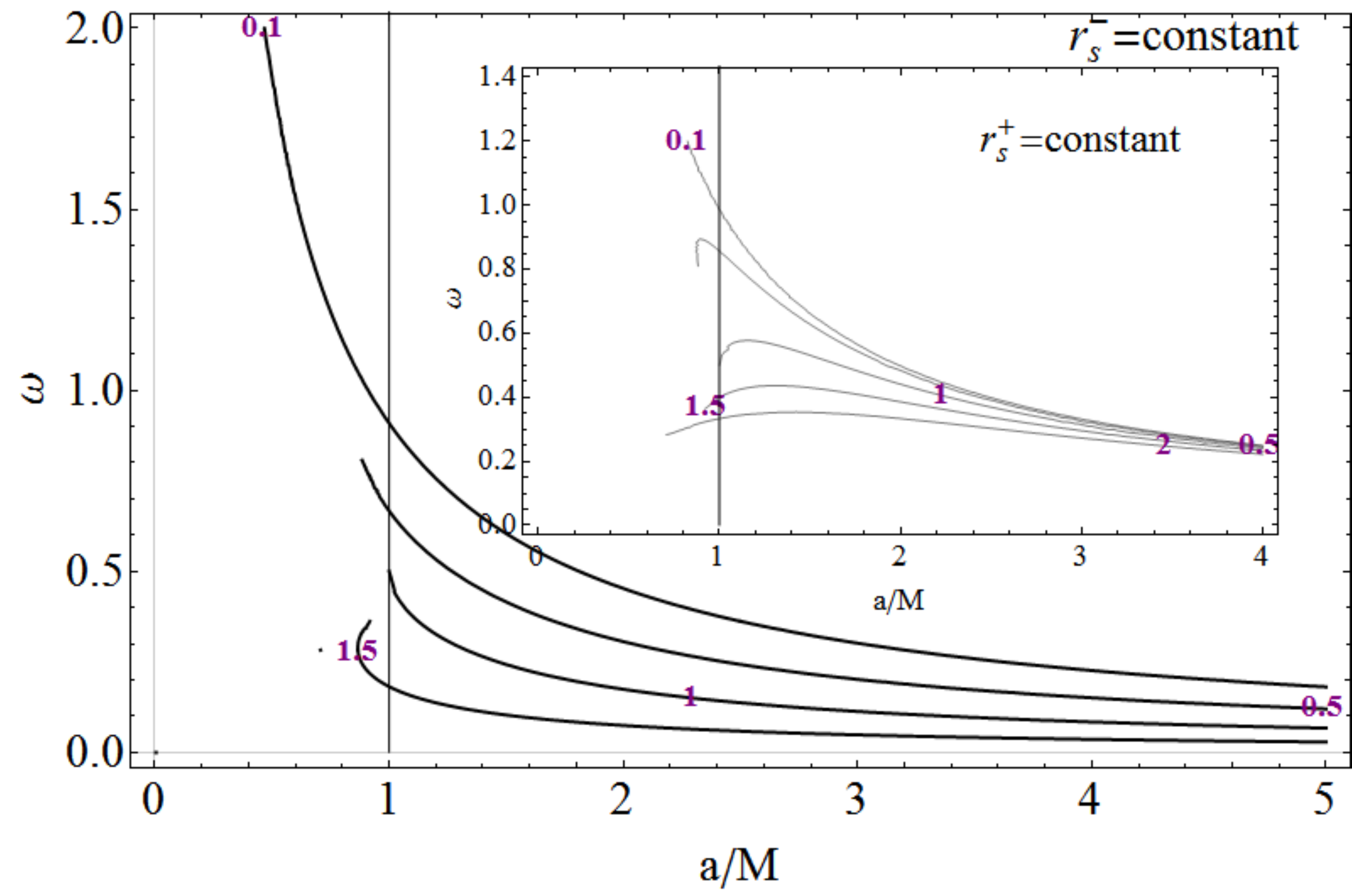}
\end{tabular}
\caption{Curves $r_s^-=$constant  and $r_s^+=$constant (inside panel) in the plane
$(\omega,a/M)$. The numbers denote the constant radii $r_s^{\pm}/M$  (light cylinders).
The angular momentum and the velocity  $(a,\omega)$ for $r_s^{\pm}(a,\omega)=0$ are related by $\omega=M/a$.
}
\label{Fig:Trav-inr}
\end{figure}

\textbf{{Zero Angular Momentum Observers}}

\begin{figure}[h!]
\centering
\begin{tabular}{c}
\includegraphics[scale=.45]{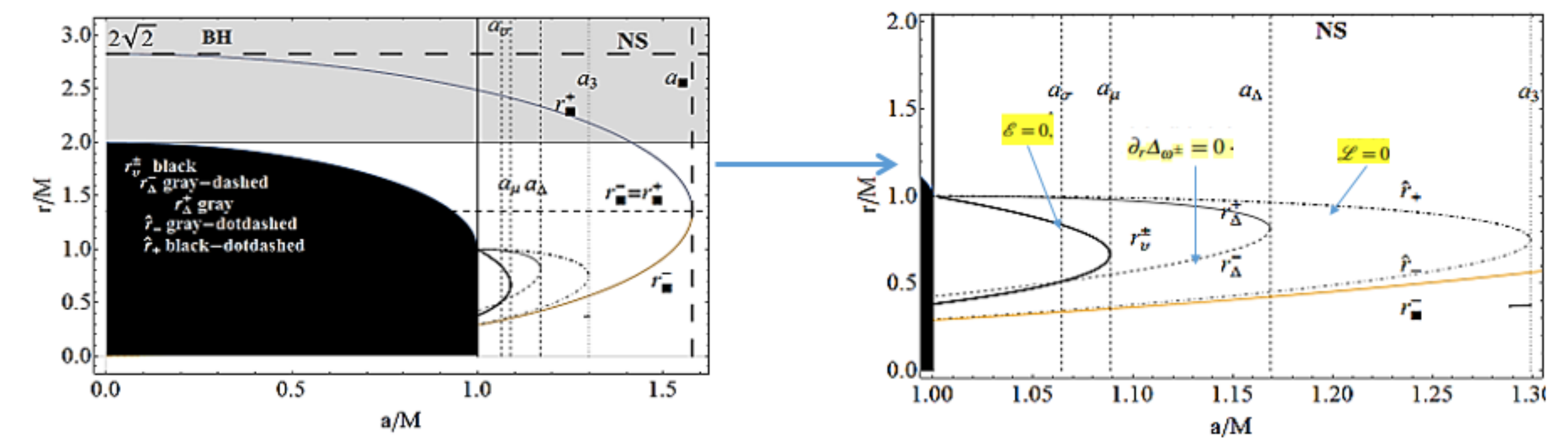}
\end{tabular}
\caption[font={footnotesize,it}]
{ The radius $r(a)$, solution of $\partial_r\Delta_{\omega_{\pm}}=0$, i.e., it represents the critical points of the separation parameter  $\Delta{\omega_{\pm}}\equiv\left.(\omega_+-\omega_-)\right|_{\pi/2}$  on the equatorial plane $\theta=\pi/2$. The radius $r_{\upsilon}^{\pm}$, where the orbital energy $\mathcal{E}=0$, and the orbits $\hat{r}_{\pm}$, for which $\mathcal{L}=0$, are also plotted. Dashed lines represent the  spins  $a_{\sigma}=1.064306M$, $a_{\mu}=4\sqrt{2/3}/3 M\approx1.08866M$,   $a_{\Delta}=1.16905M$ and ${a}_3={3 \sqrt{3}}/{4}M$.
The black region corresponds to $r<r_+$.  The radii $r_{\blacksquare}^{\pm}:\partial_{a}\Delta_{\omega_{\pm}}=0$ are plotted as functions of $a/M$--from \cite{observers}.}
\label{Fig:L0V0Zamos}
\end{figure}

\begin{figure}[h!]
\centering
\begin{tabular}{cc}
\includegraphics[scale=.5]{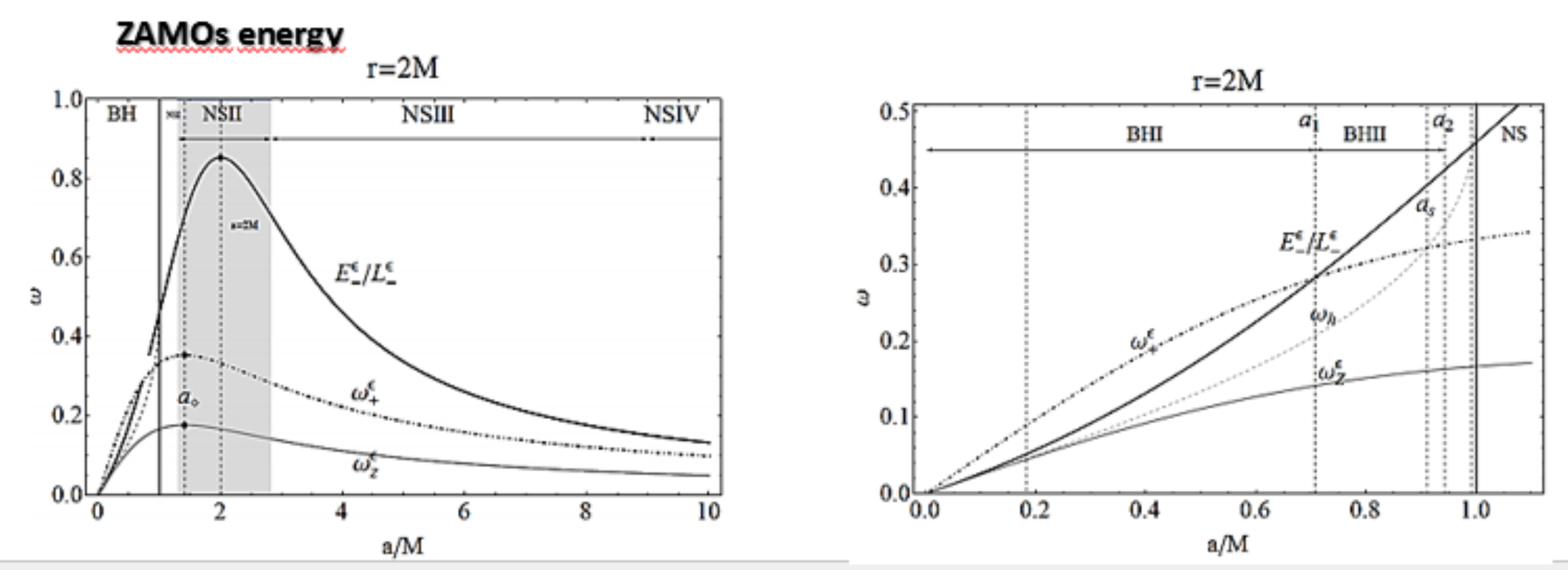}
\end{tabular}
\caption[font={footnotesize,it}]{Left panel: The ratio $\mathcal{E}^{\epsilon}_{-}/\mathcal{L}^{\epsilon}_-$ and  the angular momentum of the ZAMOs
$\omega_{Z}^{\epsilon}$ as a function of $a/M$ in the static limit $r=r_{\epsilon}^+$.
The angular momentum $\omega_+^{\epsilon}\equiv\omega_+(r_{\epsilon}^+)$ which is a boundary frequency for the stationary
observer (outer light surface) is  plotted (gray curve). The radius $r_{\epsilon}^+$  is defined by the condition
$\omega_-(r_{\epsilon}^+)=0$, $\omega_{h}$ is the ZAMOs angular velocity  on $r=r_+$, i.e.  $\omega_{\pm}(r_{\pm})=\omega_{h}$. The maxima, {related to the orbits $r_e(a)$}, are denoted by points.  A zoom of this plot in the \textbf{BH} region is in the right panel (
from \cite{observers}).}
\label{Fig:L0V0Zamos1}
\end{figure}

Some properties of spacetimes with  bottlenecks shown  in Fig.\il\ref{Fig:QPlot} can be related to the dynamics of ZAMOs, which are defined by the condition
\be
\mathcal{L}\equiv u_{\alpha}\xi_{(\phi)}^\alpha=
g_{\alpha\beta}\xi_{\phi}^{\alpha}p^{\beta}= g_{t\phi}\dot{t}+g_{\phi\phi}\dot{\phi}=0\quad
(
{d\phi}/{dt}=-{g_{\phi t}}/{g_{\phi\phi}}\equiv\omega_{Z}=(\omega_++\omega_-)/2
).
\ee
The variation of the orbital frequencies of the ZAMOs with the  spin,  for different orbits,
 shows the existence of extreme orbits as shown in Fig.\il(\ref{Fig:L0V0Zamos1}),
$ \left.\partial_a \omega_Z\right|_{\pi/2}=0,  \quad  \omega_e\equiv\omega_{Z}(r_e)$, where
\bea\label{Eq:cii-Scli}
r_e\equiv
&&
\frac{\sqrt[3]{3} a^2+\Upsilon^2}{3^{2/3}\Upsilon},\quad \Upsilon\equiv\sqrt[3]{9M a^2+\sqrt{3} \sqrt{a^4 \left(27M^2-a^2\right)}}.
\eea
The relation  between ZAMOs orbital frequency and light frequencies  $\omega_{\pm}$ can be inferred from Eq.\il(\ref{Eq:ex-ce}).
In Figs.\il\ref{Fig:L0V0Zamos} and \ref{Fig:L0V0Zamos1}  some properties of the ZAMOs dynamics are shown, which follow from
the analysis of  the Killing bottleneck and the  light surfaces, Fig.\il\ref{Fig:Trav-inr-B}.

These spins are significant for the orbital properties of the  ZAMOs which exist exclusively inside the ergoregion of
\textbf{WNS} with $a\leq 1.31M$
\footnote{The constant  $\mathcal{L}$ and $\mathcal{E}$ shown in Figs\il(\ref{Fig:L0V0Zamos})  and Figs\il(\ref{Fig:L0V0Zamos1})  are constant of motion  associated  respectively to $\xi_t$ and $\xi_{\phi}$  and compose the rotational version of the Killing fields i.e. the
canonical vector fields
$\tilde{V}\equiv(r^2+a^2)\partial_t +a\partial_{\phi}$
 and $\tilde{W}\equiv\partial_{\phi}+a \sigma^2 \partial_t$, $\sigma=\sin \theta$. The contraction of
 the geodesic four-velocity with $\tilde{W}$ leads to the non-conserved quantity
$\mathcal{L}-\mathcal{E} a \sigma^2$, which  on the equatorial plane reduces to the constant   $ \mathcal{L-E} a$.
Considering the principal null congruence
$
\gamma_{\pm}\equiv\pm\partial_r+\Delta^{-1} \tilde{V}$, there is
the angular momentum $\mathcal{L}=a \sigma^2$, that is $\bar{\ell}=1$ (and $\mathcal{E}=+1$, in proper units), every principal null geodesic is then characterized by $\bar{\ell}=1$, with
$\mathcal{L=E}=0$ on the horizon.
In this analysis, the dimensionless radius
$R\equiv r/a$ is relevant. For more discussions on the role of this ratio as $\bar{\mathcal{L}}=\mathcal{L}/a$ and $\bar{\ell}=\mathcal{L}/\mathcal{E}a$, see \cite{observers}.}.
A more detailed analysis is performed in   Figs.\il\ref{Fig:L0V0Zamos} and \ref{Fig:L0V0Zamos1}. See also
\cite{observers}. 
From the analysis of  Fig.\il\ref{Fig:L0V0Zamos} it follows that the radii connected with the
the {\emph{frequency interval}}
$
\Delta_{\omega_{\pm}}\equiv \omega_+-\omega_- $  are relevant to Killing bottlenecks:
\bea&&\nonumber
r_{\blacksquare}^+\equiv\eta  \cos \left[\frac{1}{3} \arccos \left(-\frac{8 a^2}{\eta ^3}\right)\right],
\quad r_{\blacksquare}^-\equiv\eta  \sin \left[\frac{1}{3}\arcsin\left(\frac{8 a^2}{\eta ^3}\right)\right],\quad\eta\equiv\frac{2 \sqrt{8M^2-a^2}}{\sqrt{3}},
\\\label{Eq:def-tru-all-ru}
 && \mbox{for }\quad r\in]0,2\sqrt{2}M[ \quad\mbox{
where}\quad r_{\blacksquare}^{\pm}:\, \left. \partial_a\Delta_{\omega_{\pm}}\right|_{r_{\blacksquare}^{\pm}}=0 \quad (maximum\quad points)
\eea
The plots of Fig.\ref{Fig:L0V0Zamos} show also the radii related to the frequency gap  $\Delta \omega_{\pm}$.
These radii are related to  properties generally associated with so called ``repulsive effects" of \textbf{NSs}.
In this sense, the presence of ripples in the frequency sheets seem to be a related effect.
It is also to be noted that the ripples are defined as a gap restriction in the frequency sheets; more  precisely, they
could be interpreted as due to the existence of two null orbits $r_1\leq r_2$, where $\Delta \omega_{\pm}(r_1,r_2)$ is a minimum. That is, there is a pair of points $(r_1,r_2)\in r_s\times r_s$, on the  light surface of a selected spacetime,
where the photon orbital frequencies interval (range of possible timelike orbital frequency) is minimized. Clearly,
the limiting case occurs for $r_1=r_2=r_*:\Delta \omega_{\pm}(r_*)=0$, i.e.,  in $r_*\in\{0,r_\pm\}$. In this sense,
 the horizons and singularities can be interpreted as  the limiting  cases of the Killing bottlenecks (``horizons remnants").
For more details see \cite{renmants}.
%

\section{Horizon extension: Unveiling BH--NS connections}
\label{Sec:extende}
Figure \ref{Fig:L0V0Zamos} shows a bottleneck configuration for spins $a\in ]M,1.31M]$.
We now introduce the concept of metric
bundle $g_{\omega}$ as the  collection of Kerr metrics within the  parameter range $a\in[a_0,a_g]$ with
\bea\label{Eq.lcospis}
a_{\omega}^{\pm}(r,\omega;M)\equiv\frac{2 M^2 \omega \pm\sqrt{r^2  \omega ^2 \left[M^2-r (r+2M)  \omega ^2\right]}}{(r+2M)  \omega ^2}
\eea
and a   constant  frequency $\omega$, which characterizes the bundle $g_{\omega}$. Any geometry of the  bundle possesses two distinct lightlike orbits,
$r_1\leq r_2$,  whose frequencies coincide with the characteristic frequency of the bundle, i.e.,  $\omega(r_1)=\omega(r_2)=\omega$, constrained within the limiting  geometries  with  $a_0$ and $a_g$.
Moreover, the orbital distance $(r_2-r_1)$ reaches a maximum in the bundle and is null on the borders $a_0$ and $a_g$.
\begin{figure}[h!]
\begin{center}
\begin{tabular}{cc}
\includegraphics[width=.6\columnwidth]{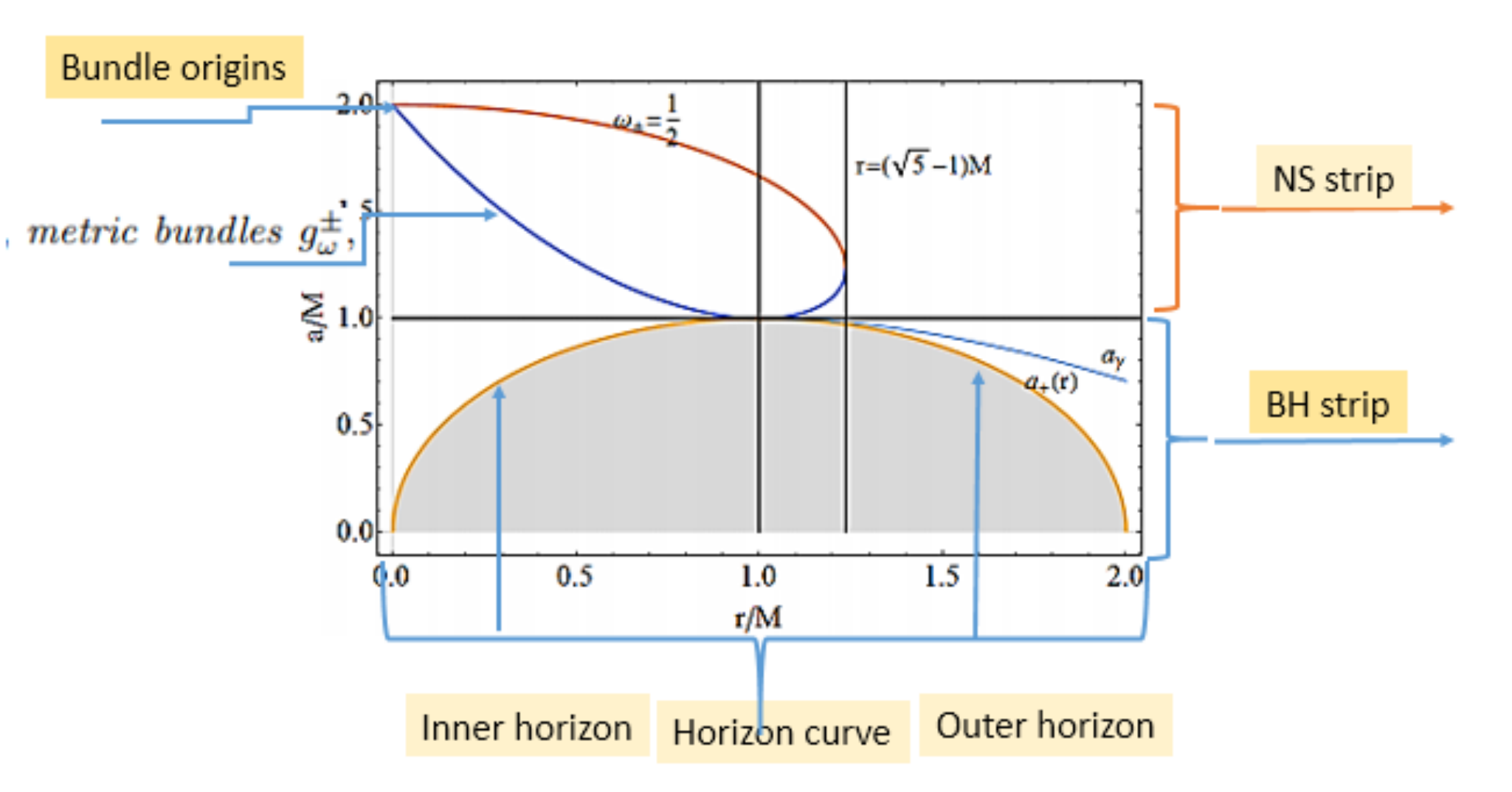}
\includegraphics[width=.55\columnwidth]{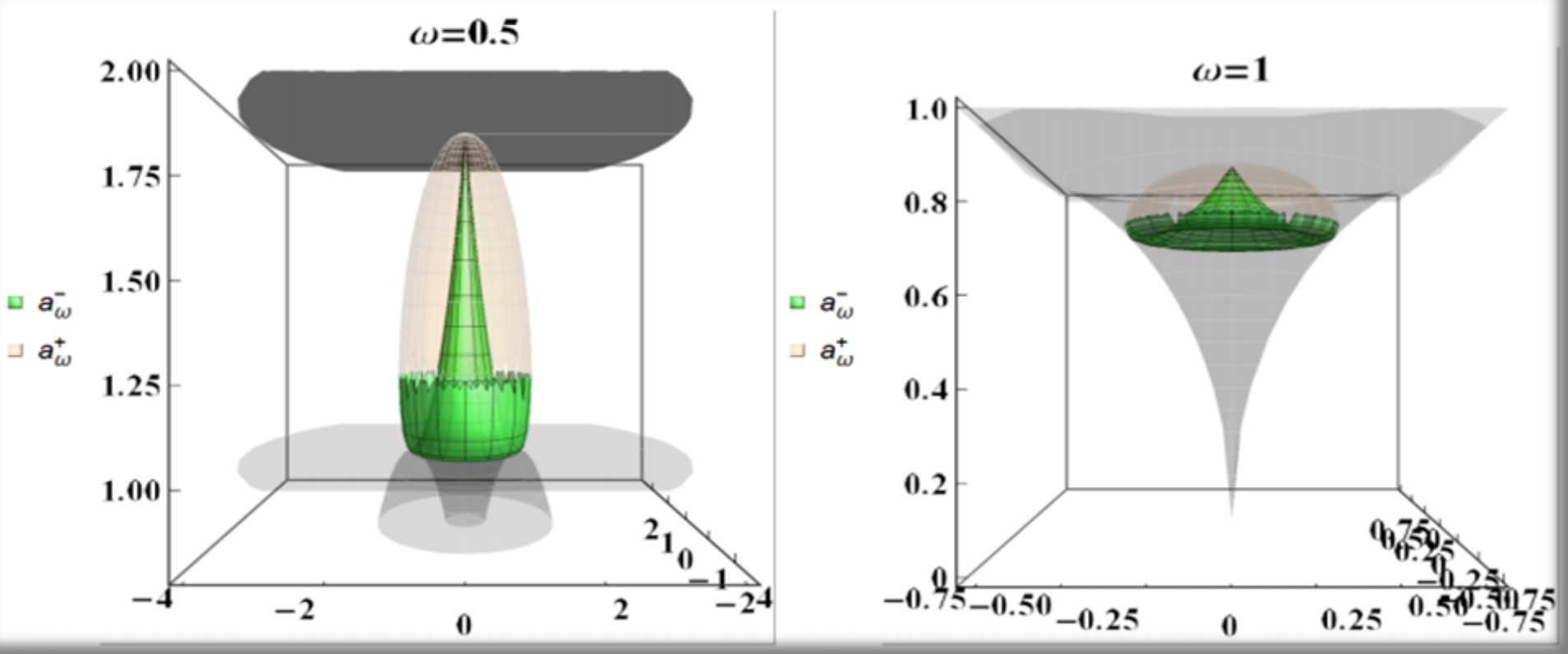}
\end{tabular}
\end{center}
\caption[font={footnotesize,it}]{The metric bundle $g_{\omega}$ for the spin  $\omega_{\pm}=\omega_h=1/2$ and bundle origin $a_{0}=2M$. The horizon curve ($a_{+}(r)\equiv\sqrt{r(2M-r)}$) and $a_{\gamma}$, solution of $r=r_{\gamma}$ where $r_{\gamma}\in \Sigma_{\epsilon}^+$  is the photon orbit in the ergoregion of the Kerr \textbf{BH}, are also plotted. The right panel shows a  confinement of the metric bundle in the region bounded by the inner horizon  and the bundle at $\omega=0.5$.}
\label{Fig:Pilosrs2}
\end{figure}
\begin{figure}[h!]
\centering
\begin{tabular}{cc}
\includegraphics[width=.7\columnwidth]{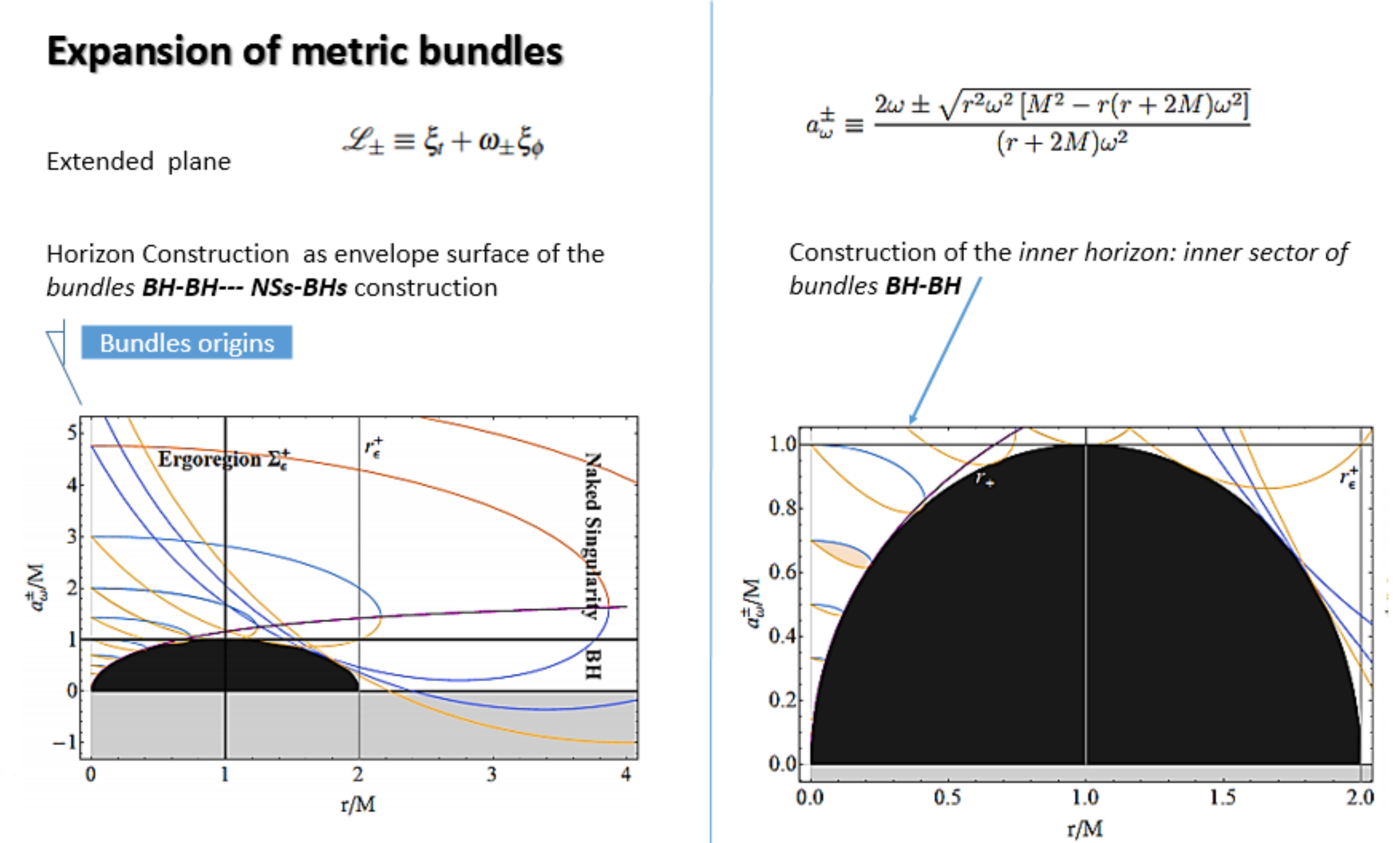}
\end{tabular}
\caption[font={footnotesize,it}]{The metric bundle $g_{\omega}$ with origins  $a_0>0$. The horizon appears as the envelope surface.}
\label{Fig:Pilosrso}
\end{figure}
\emph{All and only} the geometries of that bundle  share  this property.
From Fig.\ref{Fig:Pilosrs2} we derive some general properties of the metric bundles:
\textbf{1)} The vertical axis of the extended planes contains all  the {\emph{origins}}  $a_0\in]0,\infty[$ of the metric bundles $g_{\omega}$.
\textbf{2)} The curve $a_{+}$ represents the horizon in the extended plane $\pi^+$ and the   boundary spin $a_g\in]0,M]$ is associated
to this  curve.
\textbf{3)} We can identify a correspondence between the inner horizon $r_-$ and outer horizon $r_+$ on  the curve  $a_+$, as shown
in Fig.\il\ref{Fig:Pilosrs2};
we identify the \textbf{BH}-strip in the lower part of the panel and the \textbf{NS}-strip in the upper one as indicated.
\textbf{4)} The metric bundles $g_{\omega}$ are closed as a consequence of  the closing of the $\xi_{\phi}$ orbits.
\textbf{5)} The metric bundles $g_{\omega}$ are tangent to the horizon.
\textbf{ 6)} The metric bundle shown  in Fig.\il\ref{Fig:Pilosrs2} corresponds to the characteristic frequency  $\omega=0.5$
and origin $a_0=2M$, which is the upper limit of the spin region, where the Killing bottleneck appears.
The bundle  is tangent to the Killing horizon at the point  $r_h=M$, corresponding to the spin  $a_g=M$.
This bundle is related  also to  the light surfaces with  $\omega_0=0.5$ and $\omega_0=1$ as seen in Fig.\il\ref{Fig:Trav-inr-B}.
We note that this special metric bundle, corresponding to the spin $a_g=M$,  is regular on the point on the horizon for the extreme Kerr \textbf{BH} and relates  the spin-origin $a_0$ and the tangent point  $r_h$  on the horizon with  the spin  $a_g$ defined by the tangent point. It is clear that the horizon curve on the extended plane corresponds to a  unique  horizon frequency $\omega_H$.  It is possible to show that the correspondence $(a_0, a_g, \omega_H)$ can be set as a bijection.

Translating the bundle  origin $a_0$ in all the  range $a_0>0$ in Fig.\il\ref{Fig:Pilosrso},
some relevant properties of the metric bundles and the Killing horizon appear:
\textbf{i)} All the metric bundles are tangent to the horizon.
\textbf{ii)} The metric bundles do not penetrate the horizon.
\textbf{iii)} The space subtended by the horizon curve in the extended  plane is not described by any metric of any bundle
(in the sense of our analysis).

From the above properties, we can derive the following consequences:
\textbf{1)} The frequencies  $\omega$ of each bundle $g_{\omega}$ (and on every point $r$ of each bundle) are all and only those of
the horizon frequency  $\omega_H$. It is, therefore,  sufficient to know the horizon frequency $\omega_H$ in the extended plane to fix the photon orbital frequencies (and therefore the physical observer  frequency range) in each point of any  \textbf{BH} or \textbf{NS} geometry of the Kerr spacetime  family.
\textbf{2)} The horizon arises as the \emph{envelope surface } of all the metric bundles.
\textbf{3)} The part of the horizon curve corresponding to the inner horizon in $\pi^+$ is built partially by metric bundles  all contained in the region inside the inner  horizon, i.e.,  confined in $a_g\in [0,M]$, $r\in[0,M]$ and   $a_0\in[0,M]$.
These bundles have origins in  $a_0\in [0,M]$. However, these special bundles are not sufficient to construct, as envelope surface,
the whole inner horizon  in the extended plane.
\textbf{4)} The bundles necessary for the construction of the other part of the  the inner horizon  in $\pi^+$  have  origin in
$a_0\in[M,2M]$, i.e., in the  \textbf{WNS}  geometries, where the Killing bottleneck appear.
\textbf{5)} The Killing bottleneck appears to be related to the properties of these special metric bundles, which are
involved in the construction
 of the inner horizon as envelope surface.
\textbf{6)} The portion of the  horizon curve corresponding  to the outer  Killing horizon on the equatorial plane  is constructed by
metric bundles with origins  $a_0>2M$, which we identify as \textbf{SNS}.

We close this section noting that
  the whole  set of  photon limiting orbital frequencies $\omega_{\pm}$ (or alternatively the light surfaces $r_s^{\pm}$) of  a single Kerr geometry  with spin $\bar{a}$  (and therefore the range of orbital frequencies for the physical observers) in the extended plane,
	is given by the collection of  points of all the metric bundles on the horizontal lines $\bar{a}=$constant   in
	Figs.\ref{Fig:Pilosrs2} and \ref{Fig:Pilosrso}.
	These frequencies are all and only those of the horizon frequency $\omega_H$.
We discuss the interpretation  of this result in the next conclusive section.
Finally, we   mentioned above that the internal bundles with origin
in  $a_0\in [0,M]$  are all confined in  the region of the inner horizon in  $\pi^+$, i.e., in $a_g\in[0,M].$
This can be shown in several ways.
In particular, considering again the horizon  frequencies  $\omega_{h}^{\pm}$, we introduce the
radii $r^{\mp}_{\mp}$, defined as
\bea\label{Eq:mart-re}
&&
r^{-}_{-}=\frac{1}{2} \left(\sqrt{\frac{32 r_-}{a^2}-a^2+6 \sqrt{1-a^2}-22}-r_-\right):
 \omega_-(r_-^-)=\omega_-(r_-)=\omega_h^-
 \\
 &&
 r_+^+=\frac{1}{2} \left(\sqrt{\frac{32 r_+}{a^2}-a^2-6 \sqrt{1-a^2}-22}-r_+\right):\omega_+(r_+^+)=\omega_+(r_+)=\omega_h^+,\\
 &&\nonumber\mbox{where}\quad
 (r_-^-<r_-)<(r_+<r_+^+),
\eea
where we have used dimensionless units. The radii  $(r_+^+,r_-^-)$ correspond to  photon orbits  with
frequencies $\omega_{\pm}$ of the $\textbf{BH}$ horizons $\omega_h^{\pm}$. The orbital frequency of the inner horizon  has a replica on an orbit  $r_-^-\in[0,r_-]$- ``inner horizon frequency confinement". {This result  is in agreement with the \textbf{BHs} thermodynamic properties discussed in Sec.\il(\ref{Sec:Stationar})}.
Also, the relation $(a_g,a_0, r_h)$, where $r_h$  is obtained through $\omega_h$,  has been mentioned to be  bijective. We can show this through  the  relation between
$(a_g,a_0)$  \cite{renmants}
 \bea\label{Eq:agar}&&
\forall \; a_0>0,\quad a_g\equiv\frac{4 a_0M^2}{a_0^2+4M^2}\quad\mbox{where}\quad a_g\in[0,M]\quad \mbox{and}
\\
&& \lim_{a_0\rightarrow0}a_g=
\lim_{a_0\rightarrow\infty}a_g=0,\quad a_g(a_0=2M)=M.
\eea
This relation also allows us to formalize the
\textbf{BH-BH}  correspondence (construction of the inner horizon as an envelope surface in $\pi^+$), the   \textbf{BH-WNS} relation
 (inner horizon construction) and  the \textbf{BH-SNS}  relation (construction of the outer horizon). Particularly,  the horizon  relates in the  extended plane \textbf{BHs}  with \textbf{NSs} (through the origins of the metrics bundles $a_0$ and  $a_g$). In this sense,
the \textbf{NSs} can be interpreted as necessary for the construction (as envelope surface) of the inner and outer horizons in the extended plane.
Note that the static case of the Schwarzschild geometry, $a=0$,  can be seen  as the limiting case  in  Figs.\il\ref{Fig:Pilosrs2} and
\ref{Fig:Pilosrso}, where the horizon  $a_g=2M$  is in correspondence with the limiting \textbf{SNS}  bundle with origin  $a_0=+\infty$.
The plane $\pi^+$ in Figs\il(\ref{Fig:Pilosrs2}) and (\ref{Fig:Pilosrso}) have  several symmetries. Note that in
Fig.\il(\ref{Fig:Pilosrso}) negative values of $a_0$ and $a_g$ are possible in the metric bundles and are  related  to
$\omega_{\pm}<0$ frequencies, which are possible  outside the ergoregion ($r>r_{\epsilon}^+$). The quantity
$\mathcal{A}_{r_{\pm}}^{+}=\pi/2$ is the area of the region of $\pi^+$ bounded by the horizon
 (dimensionless quantities); $\mathcal{A}$  is the area of the regions in  the extended plane $\pi^+$ bounded by the curves $a_{\omega}^{\pm}$, defining the  metric bundles $g_{\omega}$.
$\mathcal{A}$ is a decreasing function of the frequency $\omega$, shrinking  at the  origins $a_0<M$, i.e. $\omega_0=M/a>M$, where $g_{\omega}^{\pm}$ are all bounded by the inner horizons; viceversa, the region areas  grow as the spin-mass ratio increases in the \textbf{NS} geometries.
We repeated this analysis in  the case of Reissner-Nordstr\"on and  Kerr-Newman spacetimes and similar results are found.

\section{Final remarks}
\label{Sec:remark}

In this work, we investigated the properties of stationary observers on the equatorial plane of the Kerr spacetime. The generalization to the off-equatorial case as well as the Reissner-Nordstr\"om and  Kerr-Newman spacetimes is presented elsewhere \cite{renmants}.
We focused on the behavior of the frequency of
stationary observers. To emphasize its peculiarities, we introduced the concept of Killing throats and bottlenecks. If we consider the frequency as a function of the spin, certain features appear that are better explained by introducing the concepts of extended planes and metric bundles. In the case of the Kerr metric on the equatorial plane, the extended plane is essentially equivalent to the function that relates the frequency with the spin.

Metric bundles and horizons remnants appear related to the concept of
pre-horizon regimes.
There is a pre-horizon regime
in the spacetime  when  there are  mechanical effects allowing circular orbit  observers
to recognize the close presence of an event horizon.
This concept was introduced in \cite{de-Felice1-frirdtforstati} and detailed for the Kerr geometry
in \cite{de-FeliceKerr,de-Felice-anceKerr,de-Felice first Kerr}.
The  analysis  of the pre-horizon structure
  led to the conclusion that a gyroscope would observe a memory of the static case
in the Kerr metric.
It is clear that this aspect could have an essential relevance in the investigation of the collapse \cite{de-Felice3,
de-Felice-mass,de-Felice4-overspinning}--see also \cite{Chakraborty:2016mhx}.

In the extended plane of the Kerr metric,  the frequency  on the horizons determines a set of metrics, a  metric bundle,
 describing  in general \textbf{BHs} and \textbf{NSs}, and also the limiting frequencies for stationary observers, which correspond to
light-like orbits. This fact can be interpreted as determining a connection between \textbf{BHs} and \textbf{NSs}. In the extended plane,
 the \textbf{NS} solutions  have a clear  meaning in relation to the construction as envelope surface of portions of the horizon in $\pi^+$.
  The inner \textbf{BH} horizon is connected to the origin of  \textbf{BH}  bundles and
 the outer horizon  establishes \textbf{BHs-SNSs} correspondence. Moreover, the horizon
  in the extended plane can be interpreted as the envelope surface of all metric bundles. On the other hand,
	the metric bundles are all defined by all and only the frequency
   of the horizon. In this sense, the corresponding inner horizon in $\pi^+$  is partially constructed by \textbf{BHs} metric bundles. The inner horizon is completed by bundles including \textbf{BHs} and \textbf{WNSs}.  This property appears related with the Killing bottlenecks appearing in the light surfaces. Interestingly,  the outer horizon in $\pi^+$ is  generated by \textbf{SNSs} metric bundles.   It is then possible to argue that this result could be
of interest for the investigation of the gravitational collapse. Indeed, suppose that the collapse is a quasi-stationary process in which
each state is described by a Kerr spacetime. Since rotating astrophysical compact objects are characterized by spin parameters, which correspond to \textbf{NS} configurations ($a/M>1)$, the formation of a \textbf{BH} ($a/M\leq 1)$ would necessarily imply passing through a series of states with spin parameters in the \textbf{NS} regime. In this case, the extended plane of the Kerr metric as described above could contain the different states which are necessary for the formation of a \textbf{BH}.  This fact has the interesting consequence that only  horizon frequencies   in extended plane determine   the frequencies $\omega_{\pm}$ at each point, $r$, on the equatorial plane  of a Kerr \textbf{BH} or \textbf{NS} geometry. All the frequencies $\omega_{\pm}(r)$ on the equatorial planes are only those of the horizon in
$\pi^+$. Another relevant aspect connected with this fact is the confinement of the horizon in the sense of the frequencies given in
 Eq.\il(\ref{Eq:mart-re}). These aspects are currently  under investigation.


\subsubsection*{Acknowledgments}

D.P. acknowledges support from the Junior GACR grant of the Czech Science Foundation No:16-03564Y.
This work was partially supported by UNAM-DGAPA-PAPIIT, Grant No. 111617.
{D.P. is grateful to Donato Bini, Fernando de Felice and Andrea Geralico for  discussing  many aspects of this work.}


\end{document}